# Reliable Determination of Contact Angle from the Height and Volume of Sessile Drops

Fred Behroozi*

Department of Physics, University of Northern Iowa, Cedar Falls, IA 50614, USA

Peter S. Behroozi

Department of Astronomy, University of Arizona, Tucson, AZ  85721, USA

**ABSTRACT**

Contact angle is an important parameter in characterizing the wetting properties of fluids.  The most common methods for measuring the contact angle is to measure it directly from the profile curve of a sessile drop, a method with certain inherent drawbacks. Here we describe an alternative method that uses the height and volume of a sessile drop as constraints to construct its profile by numerical integration of its two governing differential equations.  The integration yields, self consistently, the average value of the contact angle along the entire contact line as well as the footprint radius of the drop and its crown radius of curvature.  As a test case, the new method is used to obtain the contact angle of pure water on two different substrates, Teflon and Lucite.  For each substrate, four drops ranging in volume from 10 $\mu l$ to 40 $\mu l$ are used. The computed contact angles are consistent across the four different drop sizes for each substrate and are in agreement with typical literature values.

---------------------------------------------------------------

* behroozi@uni.edu

## I. INTRODUCTION

The surface wetting properties of fluids are characterized by measurement of the contact angle[1-6]. Of the several methods for measuring the contact angle[7-10], the most common is to measure it directly from the profile curve of a sessile drop[11-13]. This method is convenient since, to a good approximation, the contact angle is independent of the drop size[14] for smooth and clean substrates.

However, measurement of contact angle from the drop profile suffers from several well-known drawbacks[15-18]. Chief among these is the fact that this measurement method gives the contact angle at one point of the contact line. The contact angle, however, may differ from point to point due to variations in local surface conditions[16, 17]. In addition some uncertainty is inherent in choosing the direction of the tangent line at the contact point[18].

Historically, the first commercial apparatus for measuring contact angle from the drop profile was invented by William Zisman, who attached a specially designed microscope to a platform and used it to measure the contact angle of a drop directly from its optical image[19].

In current practice, digital cameras are used to capture the drop profile, and proprietary fitting routines are employed to obtain the contact angle from the profile curve[20-23]. Often, multiple drop images are used to increase the reliability of the measurement.

An older method for measuring the contact angle was based on the observation that the height of a sessile drop, on a horizontal substrate, approaches a maximum as its volume is increased. In 1870 the German physicist Georg Hermann Quincke (1834–1924) derived a simple but approximate equation, now known as the Quincke relation, to give the contact angle $\theta_c$ as a function of the height $h$ for large drops[24, 25],

$$\theta_c = \cos^{-1}(1 - \frac{\rho g h^2}{2\sigma}). \qquad (1)$$

In Eq. 1, $g$ stands for the acceleration of gravity, $\rho$ is density, and $\sigma$ is the surface tension. The Quincke relation was in use[26] through the early decades of the 20$^{th}$ century, but it gradually disappeared from chemistry and physics textbooks by the 1960s. We note, however, that the Quincke relation has appeared anew in a recent text[27]. Elsewhere we have reviewed the Quincke relation in its historical context along with a new derivation that exposes its limitations[28].

The profile curve of a sessile drop is governed by a second order nonlinear differential equation which is easily derived by the application of the Young-Laplace capillary equation. This equation may be reduced to two coupled first order equations that are rather more convenient to work with. Even though there are no closed-form solutions of these nonlinear equations, several attempts have been made to obtain approximate solutions for the profile curve by using various perturbation methods[29-31].

The most mathematically rigorous of these attempts[32] result in several algebraic parametric equations of the form $x(\varphi,\varepsilon)$ and $y(\varphi,\varepsilon)$ where $x$ and $y$ are the reduced coordinates, $\varphi$ is the angle of the profile curve with the horizontal, and $\varepsilon$ is the perturbation parameter. In practice, the approximate solutions have not been of much use in obtaining reliable values of the contact angle.

Here we describe a more practical and reliable alternative for determination of the contact angle from sessile drops. In Section II, we first derive the exact algebraic expression that gives the contact angle in terms of the drop parameters (crown radius of curvature, surface tension, mass, footprint radius, and height). However, this equation is not of much use in practice because measurement of the crown radius of curvature for a small drop is fraught with uncertainty. Furthermore, while the height and volume of the drop have unique values, the footprint radius may depend on the direction of measurement.

The differential equations governing the drop profile are derived directly by considering the equilibrium conditions for an infinitesimal surface belt encircling the drop. Next we describe the procedure for computing a reliable value of the contact angle by numerical integration of these equations without the need to know the crown radius of curvature. Measurement of the height and volume of the drop suffices to construct the unique profile of the drop. The integration routine also gives the drop footprint radius and its crown radius of curvature.

The main advantage of our method is that with only two unique and easily measured drop parameters, height and volume, we arrive at a much more reliable value of the contact angle— that is, its average along the entire contact line. In practice, the drop volume is conveniently measured by using a precision syringe or a sensitive scale. The drop height may be determined accurately from its digital image taken next to a fine scale or by a more sophisticated tool such as a modern cathetometer.

In Section III, we describe our procedure for measuring the volume and height of several small water drops, ranging in size from 10 µl to 40 µl, on Teflon and Lucite substrates. The computational procedure is described in Section IV, where the height and volume data are used as constraints to construct the drop profile and compute the contact angle. The integration routine is simple enough to be performed on an Excel spreadsheet. The computed values (contact angle, footprint radius, and crown radius) are given for the four drop sizes in two tables, one for each of the two substrates. For comparison, we also provide the contact angle values measured directly from drop images. Furthermore, the computed drop parameters (footprint radius and crown radius of curvature) plus the measured data (volume and height) are used in the exact drop equation to check the veracity and consistency of the numerical integration. Section V is used to discuss the results and to offer some concluding remarks.

## II. THEORETICAL BACKGROUND

### A. Equilibrium Equation

Figure 1 shows a small sessile drop of mass $M$, density $\rho$, surface tension $\sigma$, and height $h$. The crown radius of curvature is $r_0$, and the footprint radius is $R$. Consider the equilibrium conditions for the shaded upper section of the drop.

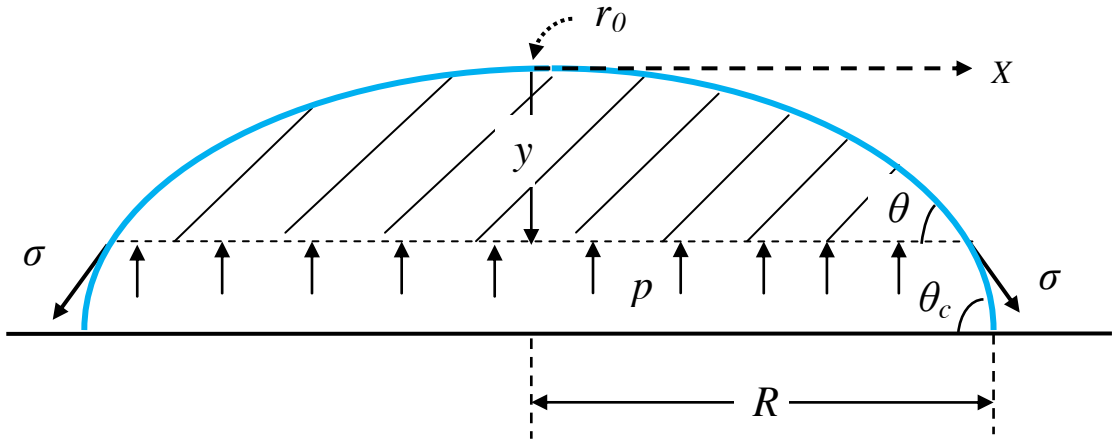

Fig. 1  The profile of a sessile drop on a surface. The hydrostatic and surface tension forces acting on the upper section of the drop (the shaded region) are shown. The gravitational force on the shaded region is not shown to avoid clutter.

The Young-Laplace relation gives the pressure difference $\Delta p$ across the surface of a liquid element in terms of its principal radii of curvature $r_1$ and $r_2$, and surface tension $\sigma$,

$$\Delta p = \sigma \left(\frac{1}{r_1} + \frac{1}{r_2}\right). \tag{2}$$

Therefore the pressure acting on the base of the shaded section in Fig. 1 is,

$$p = 2\sigma/r_0 + \rho g y, \tag{3}$$

where, the first term on the right hand side of Eq.3 is due to the differential pressure at the crown and the second term is the hydrostatic pressure. Note that the two principal radii of curvature at the apex are degenerate and equal to $r_0$.

Furthermore, if the mass of the shaded section is $m$ and its base has radius $x$, then the equilibrium of forces in the vertical direction leads to,

$$mg + (\sigma \sin\theta) 2\pi x = p\, \pi x^2. \tag{4}$$

Combining Eqs. 3 and 4, we have,

$$2\sigma/r_0 = mg/\pi x^2 + 2(\sigma \sin\theta)/x - \rho g y. \qquad (5)$$

Clearly, the shaded section of the drop may be extended down to include the entire drop in which case Eq. 5 takes the form,

$$2\sigma/r_0 = Mg/\pi R^2 + 2(\sigma \sin\theta_c)/R - \rho g h. \qquad (6)$$

Therefore,

$$\theta_c = \sin^{-1}[R/r_0 + (R\rho g h - \rho V g/\pi R)/2\sigma]. \qquad (7)$$

Equation 7 gives the contact angle $\theta_c$ in terms of the volume, height, crown radius, and footprint radius of the drop. Of these parameters, the crown radius of curvature $r_0$ is the parameter whose measurement introduces the greatest uncertainty. In practice, this measurement is nearly as prone to error as a direct measurement of the contact angle from the drop profile. Therefore Eq. 7 does not provide a practical alternative to direct measurement of the contact angle. However, it does provide a convenient check on any numerical computation as will be shown later.

**B.     Governing Differential Equations**

We note that the profile curve of a sessile drop cannot be expressed by an analytic expression[26]. However, the profile curve may be constructed for a drop of known volume and height by numerical integration of the governing differential equations of the profile. Once the drop profile is constructed the contact angle is determined.

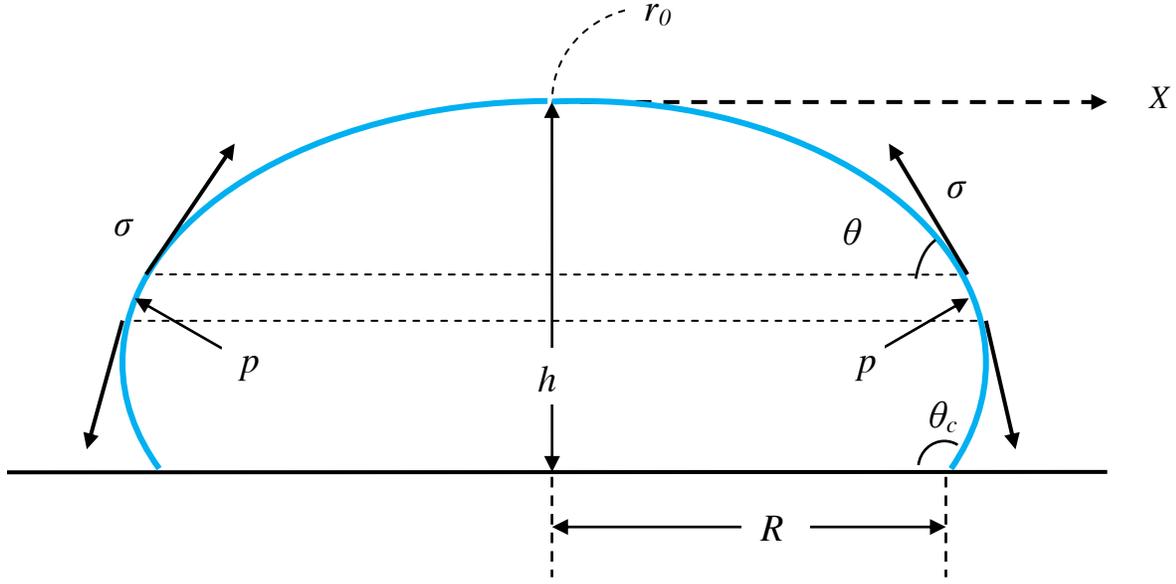

Fig. 2  The profile of a sessile drop resting on a flat surface. The crown radius of curvature is $r_0$, the height is $h$, and the footprint radius is $R$. The surface band of infinitesimal width, encircling the drop and shown between dashed lines, is in equilibrium under the action of surface tension forces and the internal pressure.

The governing differential equations are easily derived. Figure 2 shows a small drop with a crown radius of curvature $r_0$, height $h$, and footprint radius $R$. Consider a surface band of infinitesimal width and radius $x$ encircling the drop. The equilibrium of forces acting on the surface band, in the vertical direction, requires,

$$\left(\frac{2\sigma}{r_0} + \rho g y\right) 2\pi x\, dx + 2\pi x \sigma \sin\theta = 2\pi\sigma(x+dx)\sin(\theta+d\theta). \tag{8}$$

Noting that in the limit of $d\theta \to 0$,

$$\sin(\theta + d\theta) \to \sin\theta + d\theta \cos\theta, \tag{9}$$

Eq. 8 simplifies to,

$$\left(\frac{2\sigma}{r_0} + \rho g y\right) dx = \frac{\sigma}{x} \sin\theta\, dx + \sigma \cos\theta\, d\theta, \tag{10}$$

or

$$\frac{dx}{d\theta} = \frac{\sigma \cos\theta}{\left(\frac{2\sigma}{r_0}+\rho g y\right)-\frac{\sigma}{x}\sin\theta} \,. \tag{11}$$

Furthermore, noting that along the profile curve, $dy/dx = \tan\theta$, we may immediately write,

$$\frac{dy}{d\theta} = \frac{\sigma \sin\theta}{\left(\frac{2\sigma}{r_0}+\rho g y\right)-\frac{\sigma}{x}\sin\theta} \,. \tag{12}$$

Equations 11 and 12 constitute the two parametric differential equations which govern the drop profile. An auxiliary equation gives the volume of the drop,

$$V_c = \int_0^h \pi x^2 \, dy \tag{13}$$

### C. Computational Procedure

The two differential equations may be integrated numerically, starting from the drop apex ($\theta=0$), to obtain $x(\theta)$, $y(\theta)$, and the profile curve $y(x)$ when the parameters $r_o$, $\sigma$, and $\rho$ are known. However, even when $r_o$ is known, to find $\theta_c$ one must terminate the integration at the end point of the profile curve which requires knowing the value of one of the three global drop parameters (volume, height, or footprint radius).

As mentioned earlier, the value of the crown radius of curvature is not typically known. However, as described below, only two of the three global parameters suffice to construct the profile curve $y(x)$ and determine $r_0$, and $\theta_c$. For example, if the input data consist of drop height $h$, and volume $V$, one begins the numerical integration at the crown with $\theta = x = y = 0$, and with an initial choice for $r_0 = h$. The integration is stopped when $y = h$, and the computed value of the volume $V_c$ (by Eq. 13) is compared with the measured volume $V$. The initial value for $r_0$ is then adjusted up if $V_c < V$, or down if $V_c > V$ for the next round of integration. The process is repeated until $V_c \approx V$ when $y = h$ within a preset tolerance.

We note that any two of the three global parameters (volume, height, and footprint) suffice to specify the drop geometry completely. So if the volume of the drop is not known, the measured values of its height $h$ and footprint radius $R$ may be used as constraints in the numerical integration to obtain the other three parameters, $V$, $r_0$, and $\theta_c$.

When the contact angle is less than 45 degrees, the volume and footprint radius are better suited as the input data. In this case, one begins the numerical integration at the apex and proceeds with an initial choice of $r_0 = R$. The integration is stopped when $x = R$ and the computed volume $V_c$ is compared with $V$. The initial value for $r_0$ is then adjusted up if the value of $V_c < V$, and down if $V_c > V$. The process is repeated until $x \approx R$ when $V_c \approx V$ within a preset tolerance.

## III.  EXPERIMENTAL

Figure 3 shows the images of four pairs of water drops on Teflon for visual comparison. From left to right, the drop volume (in each pair), is 40 μl, 30 μl, 20 μl, and 10 μl. The drops are photographed next to a scale with fine divisions of 1/64" to show their relative heights.

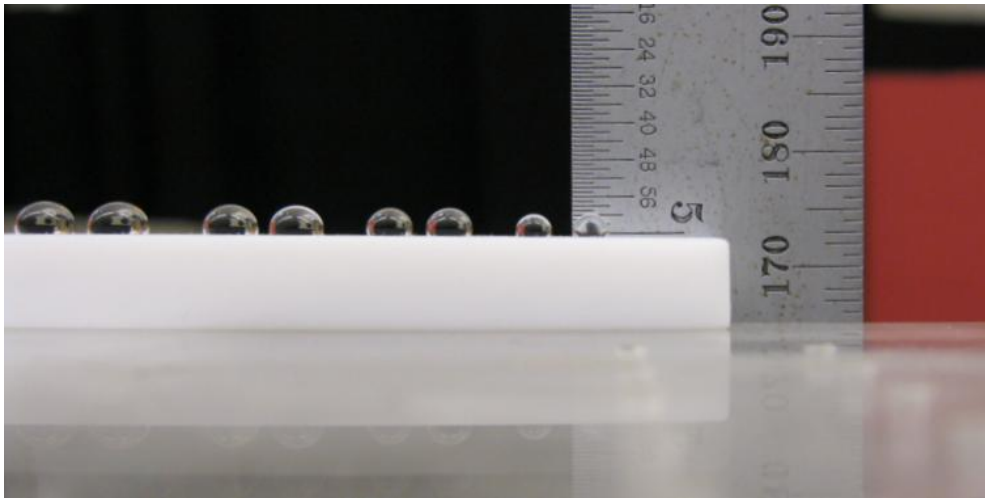

Fig. 3  Four pairs of drops are shown on a Teflon substrate. The drop volume in each pair, from left to right, are 40 μl, 30 μl, 20 μl, and 10 μl. The fine scale divisions are in 1/64 of an inch.

To measure the height of individual drops, each drop was deposited on a clean and smooth substrate (Teflon or Lucite) by a calibrated syringe and digitally photographed next to the fine scale. Before final measurements the drops were gently shaken to assume their equilibrium configurations. In our simple set up the height measurement is only accurate to $\pm 0.005\ cm$. Use of a finer scale and a higher resolution camera would enhance the accuracy of the height measurement. Our aim here is to provide typical input data for a case study of the proposed computational approach. For pure water we have used a surface tension $\sigma = 72$ dyne/cm and a density $\rho = 1$ g /cm$^3$ in our computation.

## IV.     COMPUTATIONAL REULTS

Equations 11, 12, and 13, may be integrated by any numerical technique; as explained in Appendix A, we use the midpoint method beginning at the apex and incrementing the angle $\theta$ to compute *y*, *x* and *V* stepwise along the profile curve. Note that at the starting point where $\theta = 0$, we also have *x = 0*, and *y = 0*. Therefore, referring to Eqs. 11, and 12, the values of *dy/dθ*, and *dx/dθ* are indeterminate at the apex. Employing L'Hospital's rule, we find that *dy/dθ = 0*, and *dx/dθ = $r_o$* at the apex.

In addition to the contact angle, the integration also provides, self consistently, the footprint radius of the drop (*x* at *y=h*) and the radius of curvature at its apex. The following tables list our input data (volume and height) as well as the computed values (contact angle, footprint radius, and the apex radius of curvature) for water drops, ranging in volume from 40 μl to 10 μl, on two substrates, Teflon, and Lucite.

**TEFLON:**

| Volume | Height | Crown Radius | Footprint Radius | Contact Angle |
|---|---|---|---|---|
| (*measured*) | (*measured*) | (*computed*) | (*computed*) | (*computed*) |
| (±0.1 μl) | (±0.005 cm) | (±0.005 cm) | (±0.005 cm) | (±3°) |
| 40.0 μl | 0.300 cm | 0.268 cm | 0.207 cm | 129° |
| 30.0 μl | 0.280 cm | 0.235 cm | 0.183 cm | 130° |
| 20.0 μl | 0.250 cm | 0.198 cm | 0.156 cm | 128° |
| 10.0 μl | 0.205 cm | 0.150 cm | 0.120 cm | 127° |

**LUCITE:**

| Volume | Height | Crown Radius | Footprint Radius | Contact Angle |
|---|---|---|---|---|
| (measured) | (measured) | (computed) | (computed) | (computed) |
| (±0.1 μl) | (±0.005 cm) | (±0.005 cm) | (±0.005 cm) | (±2°) |
| 40.0 μl | 0.195 cm | 0.430 cm | 0.330 cm | 69° |
| 30.0 μl | 0.180 cm | 0.375 cm | 0.297 cm | 69° |
| 20.0 μl | 0.160 cm | 0.310 cm | 0.257 cm | 69° |
| 10.0 μl | 0.130 cm | 0.235 cm | 0.203 cm | 68° |

For comparison, we give the experimental values of contact angle measured from the drop profile[33],

Water on Teflon: $\theta_c = 130° \pm 5°$

Water on Lucite: $\theta_c = 70° \pm 2°$

The computed drop parameters (footprint radius and crown radius of curvature) plus the measured data (volume and height) may be used in Eq. 7 to evaluate the contact angle. We have performed this task with the data as recorded in the above tables for all the eight drops. As expected, for each drop the contact angle evaluated from Eq. 7 is identical to the computed value within a rounding error.

## V. DISCUSSION

The new method offers a convenient and reliable technique for measuring the contact angle. The convenience is due to the relative ease of measuring the height and volume of a drop compared with the problems encountered in measuring the contact angle directly. The reliability stems from the fact that the equilibrium condition for the drop as expressed in Eq. 7, shows the intimate relation of the global parameters ($h$, $V$, and $R$) with the intrinsic ones ($\sigma, \rho,$ and $\theta_c$). The primary limit to reliability is the measurement error on the drop height. For the range of drops considered here, we find that a relative error $\varepsilon$ on the drop height translates to relative errors of $1.3\,\varepsilon - 1.7\,\varepsilon$ on the contact angle. With a modern digital camera (3-4k pixels per dimension), a macro lens, and software to correct residual lens distortions, relative errors of <0.5% in the contact angle are possible. The volume accuracy gives a secondary limit on reliability; for fluids that can evaporate, the use of a sensitive scale may help reduce systematic biases.

The computational method outlined here uses as input two of the three global parameters of the drop. So the resulting value of the contact angle is a global average and reflects the condition along the entire contact line of the liquid /solid interface. Consequently, local rough spots or surface defects do not significantly affect the outcome.

For large drops, the height approaches a maximum value asymptotically as the volume is increased, rendering the height data insensitive to volume changes. Similarly for small contact angles, measurement of the drop height suffers from large relative uncertainty. In both of these cases, the volume and footprint radius should be used as input data for the computation. Indeed, a simple, useful extension of our method would be to couple it with a Markov Chain Monte Carlo algorithm so that all three quantities (volume, footprint radius, and height) could be used simultaneously to constrain the range of allowed contact angles.

We mention in passing that when three global parameters are available as input data in the numerical integration one can obtain both the contact angle $\theta_c$ as well as the surface tension $\sigma$. In practice, the volume $V$ and height $h$ of a sessile drop can be measured reliably and thus constitute two of the three measured global parameters. When the contact angle is greater than 90°, the equatorial radius $R_e$ is the best choice for the third parameter. And when the contact angle is less than 90°, the footprint radius $R$ is used as the third parameter. In either case, integration of the governing differential equations with trial values of the apex radius of curvature $r_o$ and surface tension $\sigma$ generate an initial theoretical profile curve. The difference between the theoretical volume, height, and equatorial radius (or footprint radius) with their experimental counterparts is then used to update the trial values of surface tension and apex radius of curvature. By using a two-dimensional fitting routine the procedure is iterated until the best fitting values for the apex radius of curvature and surface tension are found which in turn provide the contact angle. However, since the footprint radius of a drop varies from point to point along the contact line due to local surface defects, its measured value is necessarily uncertain. Consequently, the computed values of surface tension and contact angle by this method reflect this uncertainty.

In contrast, when the surface tension is known, determining the contact angle from the height and volume of small drops is relatively simple and free from most of the errors and uncertainties associated with direct measurement of the contact angle from the drop image. Indeed, when the volume and height of a sessile drop is measured accurately, the method described in this paper yields a very reliable value of the contact angle because the computed value reflects the global average of the contact angle along the entire solid/liquid/vapor contact line.

**APPENDIX A: CODE IMPLEMENTATION**

Here, we describe the code used to solve Eqs. 11-13 for the contact angle given the drop height and volume. For a given guess for the crown radius of curvature ($r_0$), we can integrate Eqs. 11 and 12 using the midpoint method starting from $\theta=0$. E.g., for $x(\theta)$:

$$x(\theta + \Delta\theta) \approx x(\theta) + \Delta\theta \cdot x'\left(\theta + \frac{\Delta\theta}{2}\right)$$

where evaluating $x'\left(\theta + \frac{\Delta\theta}{2}\right)$ requires approximating $x\left(\theta + \frac{\Delta\theta}{2}\right)$ and $y\left(\theta + \frac{\Delta\theta}{2}\right)$:

$$x\left(\theta + \frac{\Delta\theta}{2}\right) \approx x(\theta) + \frac{\Delta\theta}{2} \cdot x'(\theta)$$

$$y\left(\theta + \frac{\Delta\theta}{2}\right) \approx y(\theta) + \frac{\Delta\theta}{2} \cdot y'(\theta)$$

An identical operation is applied simultaneously to calculate $y(\theta)$. The boundary conditions are that $x(0)=y(0)=0$, and, as noted in Section III, $x'(0) = r_0$ and $y'(0) = 0$. The volume $V(\theta)$ is then calculated using the midpoint method applied to Eq. 13. The integration is terminated when $y(\theta)$ reaches the drop height $h$; the angle at which this happens is by definition the contact angle $\theta_c$. This yields a method to calculate the drop volume as a function of $r_0$ and $h$, as well as the contact angle as a function of $r_0$ and $h$. We then use the Newton-Rhapson method to solve for the value of $r_0$ (if it exists) that satisfies the equation

$$V(r_0, h) - V_{measured} = 0$$

Finally, this value of $r_0$ is used as above to determine the implied contact angle $\theta_c$. The footprint radius $R$—i.e., $x(\theta_c)$—may then be used to verify the solution via Eq. 7. Even with $\Delta\theta = 10^{-4}$ rad, this process takes <0.01 sec on a modern computer. Our implementation is available at: https://bitbucket.org/pbehroozi/contactangle/src .